\documentclass[coll,preprint]{imsart}

\usepackage{amsthm,amsmath,amsfonts,natbib,graphicx}
\usepackage{url}
\RequirePackage[colorlinks,citecolor=blue,urlcolor=blue]{hyperref}
\usepackage{placeins}

\arxiv{arXiv:1206.4766v1}
\startlocaldefs

\theoremstyle{remark}

\newfont{\msbm}{msbm10 at 11pt}

\endlocaldefs
\begin{document}

\begin{frontmatter}

\title{A CB (corporate bond) pricing probabilities and recovery rates model for deriving default probabilities and recovery rates}
\runtitle{Corporate bond model}

\begin{aug}
\author{\fnms{Takeaki} \snm{Kariya}\thanksref{a}\corref{}\ead[label=e1]{kariya@kisc.meiji.ac.jp}}
\address[a]{Graduate School of Global Business \\ Meiji University
 \printead{e1}}
\affiliation{Meiji University}

\thankstext{t1}{The author is honored to contribute this paper to this
  volume for Professor Morris L. Eaton as his former student.}

\runauthor{Kariya}

\end{aug}

\begin{abstract}
  In this paper we formulate a corporate bond (CB) pricing model for
  deriving the term structure of default probabilities (TSDP) and the
  recovery rate (RR) for each pair of industry factor and credit
  rating grade, and these derived TSDP and RR are regarded as what
  investors imply in forming CB prices in the market at each time. A
  unique feature of this formulation is that the model allows each
  firm to run several business lines corresponding to some industry
  categories, which is typical in reality. In fact, treating all the
  cross-sectional CB prices simultaneously under a credit correlation
  structure at each time makes it possible to sort out the overlapping
  business lines of the firms which issued CBs and to extract the
  TSDPs for each pair of individual industry factor and rating grade
  together with the RRs. The result is applied to a valuation of CDS
  (credit default swap) and a loan portfolio management in banking
  business.

\end{abstract}



\end{frontmatter}

\section{Introduction}
In the financial industry credit risk is a central theme of interest
in banking, investment and derivatives. In the past financial economic
theory and no-arbitrage theory in mathematical finance has provided
many important results and thoughts for looking into many
credit-risk-related problems. These normative theories are very
important as a benchmark for those in financial industry. However, a
blind application of the theories to the real world may cause a big
problem, which happened in the 2008 Financial Crisis. In fact, rating
agencies wrongly applied the theories to valuation of mortgage-backed
securities and expanded the serious Subprime Problem, while insurance
companies issued credit risk swaps beyond their capacity of possibly
taking risks and made a serious impact on the US economy, which in
turn has been badly affecting world economies. In this paper, the
author often emphasizes the importance of testing the empirical
validity of ``theories'' and the limitation of the framework of
no-arbitrage theory from a practical viewpoint below.  But he never
intends to deny the importance and value of theory itself, and thinks
it important to now discuss openly about what caused the crisis among
academics and practitioners in order not to let it happen again.

In credit risk analysis in mathematical finance, a time-continuous
model is usually formulated together with no-arbitrage concept and it
is often the case that a spot rate process is used for instantaneous
interest rates, the default-event generation process is associated
with such a model as hazard model or Merton-type model and the
instantaneous recovery rate at each moment is constant. The resulting
model is usually Markovian and univariately stochastic, it
individually evaluates the default probability of each product or each
firm and the data used in applications is mostly past default rates,
rating scores, and firms’ financial data together with stock
prices. It should be pointed out that the Markovian properties, which
are the results of the assumption of time-continuous diffusion model,
do not hold for credit risk processes in reality in general. It is
because business cycles that are closely associated with credit
variations of firms are not Markovian, which is a limitation of the
model. On the other hand, \citet{kariya:2004a} treated a time-discrete
theory to include non-Markovian models.

In addition, since credit risk in reality is related to many factors
such as industry business cycles which may depend on import-export
relations, exchange rates, resource prices, etc, it does not admit a
complete model except for the case in its theoretical assumption,
meaning that the no-arbitrage concept in mathematical finance is not
necessarily effective in reality. This in turn implies that in
practice there is no unique valuation via the no-arbitrage.

In this paper we formulate a corporate bond (CB hereafter) pricing
model that enables us to derive the term structures of default
probabilities (TSDPs hereafter) and the recovery rates (RRs hereafter)
consistently with rating factors and industry factors, where the TSDP
is the set of default probabilities that a firm gets defaulted by a
future time $s$ where $s$ belongs to a positive interval from 0. Our
model is a cross-sectional empirical model valuing all the given CB
prices simultaneously at each time.

In this formulation we take into account the fact that each enterprise
has a portfolio of multiple business lines corresponding to some
different industries, and we explicitly incorporate the business
portfolio structures of individual firms together with rating factors
into the model in order to derive the TSDPs and RRs implied in CB
prices. In our case the portfolio ratios of business lines are
measured by the sales ratios of industry-wise business lines. Hence in
estimation we need the sales and rating data together with the CB
price data and bond attributes (coupon and maturity period).

Our modeling is based on three fundamental assumptions or
viewpoints. The first one is about the information content contained
in CB prices. The CB prices are assumed to be efficiently formed in
the market at each time and so reflect or contain the investors' views
on the term structure of future default probabilities for each CB over
its maturity period. In fact, their investment decision makings are
usually based on sufficient information and analysis on firms and
hence almost all the CB prices should be supposedly consistent with
the investors' views. Consequently the TSDPs we aim to derive for each
firm from all the current CB prices are regarded as the investor's
forward-looking TSDPs.  

Second, it is viewed in our that the credit condition or credit
quality of each firm in general depends on its future cash (profit)
flows, which in turn depend on the business portfolio
structure. Therefore, the credit condition or credit quality of each
firm in general depends on economic trends or business cycles that are
often different industry-wise.

The third point is related to the fact that the specification of
default correlation is very crucial in credit risk analysis. In our CB
price modeling, the default correlations are naturally introduced from
the model structure through considering those of stochastic discount
functions and the cash flow structure of CBs with defaults. On the
other hand, in time-continuous setting, the correlations are often
assumed to be constant though they in fact change constantly and
increase in the phase of downturn economy where industry-wise business
cycles are relevant. Concerning modeling the stochastic correlations,
in discrete time series analysis \citet{engle:2009} extensively treats
multivariate return processes with conditional stochastic volatilities
and conditional stochastic correlations. As terminology, when a credit
analysis contains the credit risks of multiple firms, it is often
referred to as multi-name case where the correlation structure needs
to be specified. Otherwise it is referred to as single-name case.

In applications, the implied TSDP and the implied RR that we derive
from the formulation of our CB pricing model can be applied to price
such credit derivatives as CDS (credit default swap). In fact, in this
paper we also give a formula for valuing a CDS of CB in our
discrete-time approach. Since our TSDP depends on the business
portfolio structure of the issuer, so does the pricing formula of a
CDS, though the form of the formula itself is rather well known.

There is a vast literature in the area of credit risk. Most recent
researches take a time-continuous setting. The books by
\citet{duffie:2003}, \citet{lando:2004}, and \citet{mcneil:2005} are
well known.  Most of the articles treat a single-name (univariate)
case though some recent works consider a multi-name case, e.g,
\citet{filipovic:2009}. But most of the papers do not take into
account the feedback structure of economies and simply assume
exogenous processes for credit movements without considering business
cycles. \citet{duffie:1996} viewed interest rate processes as
dependent and used a state space model though they assumed a single
Markovian process for the univariate state variable. Some papers
consider a specific factor for each firm, but the specific factors are
simply treated as some exogenous processes independently of a common
factor.

In our modeling the stochastic behaviors of the CB prices correspond
to those of the attribute-dependent discount functions, which are
formally expressed with attribute- specific forward rates. In
association with this viewpoint, \citet{collin:2001}
considers the term structure of default premia in the swap and LIBOR
markets.  \citet{feldhutter:2007} decomposes swap rate into a
common swap rate and swap spreads that are specific to credit or
counter party risk, where the spreads are referred to as convenience
yields.

The organization of this paper is as follows. In Section 2, some
important problems are discussed in the formulation of credit risk
model to be used in practice. In particular, some differences between
time-continuous models and our model are discussed from a practical
view point. In a line with the arguments in Section 2, Section 3
presents an extension of the government bond (GB) model that
\citet{kariya:1994} and \citet{kariya:1996}  proposed, where the model
is shown to have a one-month ahead predictive power for Japanese
Government bond prices. And we discuss about the relation between the
stochastic discount functions and spot and forward interest rates
which are both attribute-dependent. In Section 4 the non-defaultable
bond model is specified in detail for empirical work and an estimation
procedure is proposed. The important parts are the specifications of
the mean discount function and covariance structure of the bond
prices, which separates our model from the other models. In this paper
GB and non-defaultable bond are assumed to be synonymous.

In Section 5, we formulate the CB pricing model that enables us to
derive the implied TSDPs and RRs. This formulation is quite different
from the models in the literature in that in our modeling the concept
of the investors' forward-looking views on the TSDPs is introduced and
implemented into the model. In Section 6, we apply our results to
credit risk management in banking and pricing credit default swap
(CDS) among others.

\section{Problems in credit risk analysis}
In Section 1, concerning credit risk models developed in mathematical
finance, we discussed about some features of the models from a
practical viewpoint. Those features come from the continuous-time
setting; Markovian, univariate, instantaneous recovery rate, etc. In
addition, the following points should be well considered in the
formulation: 
\begin{enumerate}
\item[(1)] Data source to be used for credit-risk modeling--past data or current data; 
\item[(2)] Default correlation problem--univariate
(single-name) model or multivariate (multi-name) model; 
\item[(3)] Conditional model or unconditional model; and 
\item[(4)] Information on industry factors and rating factors.
\end{enumerate}
The point (1) is important for practical effectiveness of a model so
long as the model is used for a forward-looking decision-making or
investment. In fact, economic, financial and technological
environments surrounding firms are evolving constantly and rather
quickly and hence models using past data on defaults and non-defaults
and past financial data of firms do not necessarily give us a
forward-looking information to get a future credit perspective for
each firm. The model based on past data may tend to deliver a rather
backward-looking information, which may be the case of Markovian
transition model or hazard rate model, where data on defaults and
non-defaults over some past period is often used. On this point, it is
noted that each default event in the past is very firm-specific in its
nature, and defaulted firms do not exist any longer.  Credit risk
models that use current market price data on existing firms are more
forward-looking as the investors try to be rational and analytical to
make gains. Hence the CB prices that are formed in the market are
supposed to reflect the views of investors. In this sense our TSDPs
whose information is based on current cross-sectional data of CB
prices can be regarded as the TSDP of investors' perspective views and
so it will be more practically effective for decision makings.

Concerning (2) and (3), note that credit risk factors in
credit-related instruments and derivative products are generally
significantly correlated through industry factors and business cycles
and so credit correlations should be well modeled consistently in
valuation. In addition, business cycles in each industry are often
different, differently affecting firms that have different portfolios
of business lines with some overlapping for each other. Furthermore
default processes in practice affected by such business cycles is in
general non-Markovian and mutually dependent. In the literature a
Merton-type model, which is basically univariate Markovian
(conditional) geometric Brown model for stock (firm value) prices, is
often extended to a multi-name case where the correlations may be
treated as constants with copula functions. It is noted that it is not
easy to derive the TSDP from stock prices as stock does not have a
finite time horizon.

Considering (4), note that the future cash flow (profit) structure of
an enterprise depends on the portfolio structure of business lines
associated with industry factors.  The cash flow structure matters
mostly for the credit quality because most of defaults occur due to
the lack of liquidity or cash, not through the imbalances in the
financial balance sheets of firms. Since the industry factors make a
correlation structure among different credit risks of firms to a great
extent, the portfolio structure of business lines associated with
industries should be taken into account for valuing CBs, CDSs and some
other credit-related products though the concept of industry is
relative and needs to be defined in advance.

Furthermore it is remarked that in reality the recovery rate (RR)
after default is determined after a long procedure of asset evaluation
and negotiation process among those of interests, which is costly and
very time-consuming. In our formulation it is derived as the RR that
investors expect at current time and recovery is assumed to takes
place at the next coupon-paying time, as will be discussed in Section
5.

\section{Pricing non-defaultable bonds}
Suppose that there are $G$ government bonds (GBs) or equivalently
non-defaultable bonds whose prices are denoted by $P_{g}$ ($g=1,\ldots, G)$. Let $t=0$ denote the present time and let
\begin{equation}
\label{3.1}
s_{g1} < s_{g2} < \cdots < s_{gM(g)} \hspace*{10mm} g = 1, \ldots,G
\end{equation}
denote the future time points at which the $g$th bond generates the
cash flows (coupons or principal) in view of $t=0$.  These $s_{gj}$
values are measured in years, where $s_{gM(g)}$ is the maturity period. In this section and the next section future time points are
measured continuously in years. Assume that the face value is 100 (yen or dollar) and
let $c_{g}$ be its coupon rate (yen or dollar). If coupons are paid biannually, the cash flow
function $C_{g}(s)$ of the $g$th bond is expressed as
\begin{equation}
\label{3.2}
C_{g}(s) = 
\begin{cases}
0.5 c_{g} & s=s_{gm}\, , m \neq M(g)\\
100 + 0.5 c_{g} & s = s_{gM(g)} \\
0 & s \neq s_{s_{gm}}
\end{cases} \; .
\end{equation}
However, in our argument the cash flow function can be arbitrary, so
long as the future cash flows and their time points are given in
advance.

Let $D_{g}(s)$ be the attribute-dependent stochastic discount function
of the $g$th bond defined on $0 < s \le s_{aM(a)}$ with $s_{aM(a)} =
\max_{g} s_{gM(g)}$, where the whole values of $D_{g}(s)$ are realized
all at $t=0$ and $D_{g}(s)$ discounts cash flow $C_{g}(s)$ by
$C_{g}(s) D_{g}(s)$. Under these notations, our basic formulation for
modeling $G$ bond prices simultaneously at $t=0$ is based on the
following expression:
\begin{equation}
\label{3.3}
P_{g} = \sum_{j=1}^{M(g)} C_{g}(s_{gj}) D_{g}(s_{gj}) \hspace*{10mm} g=1,\ldots,G\;.
\end{equation}
This is an unconditional cross-sectional expression. In \eqref{3.3}, we
regard the realization of price $P_{g}$ as equivalent to the
realization of the whole function $\{D_{g}(s)\, : \, 0 \le s \le
s_{gM(g)}\}$ with $g=1,\ldots,G$. Hence the realizations of $G$ bond
prices correspond to those of $D_{g}{s_{am}}$,
($m=1,\ldots,s_{aM(a)}$; $g=1,\ldots,G$) and the correlation structure
of these stochastic discount functions implies those of prices.

In the spot rate approach in mathematical finance a non-defaultable
bond price at $t = 0$ is specified as the conditional expectation with
attribute-free discount function given the past and present
information, which is expressed as follows; for each individual price
\begin{equation}
\label{3.4}
P_{g}(1) = \sum_{j=1}^{M(g)} C_{g}(s_{gj}) \bar{\bar{D}}(s_{gj})
\end{equation}
with
\[ 
\bar{\bar{D}}(s_{gj}) = E_{0} \left[\exp\left( - \int_{0}^{s_{gj}} r_{u} \, du \right) \right] \equiv H(r_{0}, s_{gj}, \theta) \, ,
\]
where $\{r_{u} \, : \, 0 \le u \le s_{aM(a)}\}$ is a process of
instantaneous spot interest rates $\{r_{u}\}$ that is common to all
the bonds and $E_{0}[~]$ denotes the conditional expectation given
$r_{0}$ at 0 with respect to a risk neutral measure. But the measure
is not uniquely identified.  Here $\theta$ denotes a set of possible
parameters when a specific model such as CIR (Cox-Ingersoll-Ross)
model or Vasicek model is used for the spot rate process $\{r_{u}\}$.
In mathematical finance the conditional expectation can be regarded as
being taken under a risk neutral measure and \eqref{3.4} is claimed to hold
a.s. for all the bonds in its no-arbitrary theory. In fact,
conditioning variable $r_{0}$ is the only random variable making all
the $G$ bond prices realized. Hence in reality it does not follow that
\eqref{3.4} holds a.s. as the bond prices are not functionally related and
in fact maturities and coupon rates are different. In such a spot rate
approach, modeling the spot rate process yields the conditional
discount function through which the zero yield curve $\{ R_{u} \, : \,
0 \le u \le s_{aM(a)} \}$ defined by $H(r_{0}, s, \theta) =
\exp(-R_{s} s )$ or equivalently $R_{s} = (-1/s) \log H(r_{0},s,
\theta)$ is obtained.  In the sequel we use the real measure that
generates real data.

Another remark on this approach is that this specification of spot
rate process for bond-pricing ignores such bond attributes as coupon
rate or maturity. Empirically speaking, it is often observed that bond
prices formed in the market depend on such attributes, in which case
it is required to take into account such attribute-dependency in the
specification of the spot rate process. In modeling a swap rate
process \citet{collin:2001} and \citet{feldhutter:2007} take the
dependence of swap rates on credit attributes into account and specify
a swap rate process as the sum of an abstract risk-free rate process
$\{x_{1s} \}$ and a convenience yield process $\{x_{2gs}\}$ where they
are assumed to be independent. Here the convenience yield represents
such attributes as liquidity premium, credit premium (collateral
condition), etc.;
\[
r_{gs} = x_{1s} + x_{2gs} \; .
\]
The attribute-dependent convenience process $\{ x_{2gs} \}$ can play
an adjusting factor for fitting the model as it can be arbitrarily
specified. Using this process the discount function in \eqref{3.4}
becomes attribute-dependent;
\[ 
\bar{\bar{D}}_{g}(s_{gj}) = E_{0} \left[ \exp\left(- \int_{0}^{s_{gj}}
    r_{gs} \, ds \right)\right] \; .
\]
In this paper we do not use this spot rate approach in \eqref{3.4} with
conditional expectation but take a forward rate approach with
unconditional expression, and make it the following
attribute-dependent model as in \eqref{3.3};
\begin{equation}
\label{3.5}
P_{g} = \sum_{j=1}^{M(g)} C_{g}(s_{gj}) D_{g}(s_{gj}) \hspace*{6mm} \text{ with } \hspace*{6mm} D_{g}(s) = \exp\left( - \int_{0}^{s} f_{gu} \, du \right) \, ,
\end{equation}
where $\{f_{gs} \, : \, 0 \le s \le s_{aM(a)} \}$ is an instantaneous
forward rate term structure process whose values are realized all at
$t=0$. In other words, for each $g$ a realization of $P_{g}$
corresponds to that of the whole path $\{f_{gs} \, : \, 0 \le s \le
s_{aM(a)} \}$. Note that \eqref{3.5} is equivalent to \eqref{3.3}.

On the other hand, as an attribute-free forward rate process one may
use the time-continuous HJM (\citet{heath:1992}) model, which
describes for each individual $g$ a process of term structures
$\{f_{ts} \, : \, 0 \le s \le s_{taM(a)}\, , \, t \ge 0 \}$ with the
discount function $D_{t}^{*}(s_{gj}) = \exp\left( - \int_{0}^{s_{gj}}
  f_{ts} \, ds \right)$ attribute-free. The HJM model is specified
conditionally and the Markovian expression is usually explored with
the no-arbitrage argument. Even in this case, one single path $\{f_{s}
\, : \, 0 \le s \le s_{aM(a)} \}$ does not make \eqref{3.5} hold
a.s. for all $g$ either and so we use \eqref{3.5}.

Finally it is remarked that a specific cash flow pattern guaranteed by
holding some GBs will be a big value to such institutional investors
as pension funds or life insurance companies because they need to
match cash inflow with cash outflow over a long time horizon. In other
words, coupon and maturity are important attributes which affect
investment decisions with a future perspective. For example, depending
on cash inflow-outflow structures of investors and on future
perspectives on movements of interest rates, it may happen that a GB
of 2 year maturity and 5\% coupon is less preferred to a GB of 6 year
maturity and 3\% coupon. It is noted that those institutional
investors do not necessarily prefer de-coupon or stripped bonds
engineered by investment banking as making a portfolio from these
stripped bonds to match the cash inflows and outflows is costly and
involves additional credit risk of investment bankers.

\section{GB pricing model}
In this section, we implement an attribute-dependent GB pricing model
with a stochastic discount function $D_{g}(s)$ in \eqref{3.3} or
\eqref{3.5}. First, let $D_{g}(s)$ be decomposed into the mean function
and the stochastic deviation function as
\begin{equation}
\label{4.1}
D_{g}(s) = \bar{D}_{g}(s) + \Delta_{g}(s) \; .
\end{equation}
Substituting this into \eqref{3.5}, it follows that
\begin{equation}
\label{4.2}
P_{g} = \sum_{m=1}^{M(g)} C_{g}(s_{gm}) \bar{D}_{g}(s_{gm}) + \eta_{g} \, \hspace*{10mm} \, \eta_{g} = C'_{g} \Delta_{g} = \sum_{m=1}^{M(g)} C_{g}(s_{gm}) \Delta_{g}(s_{gm}) \, ,
\end{equation}
where
\begin{equation}
\label{4.2a}
C_{g} = \left(C_{g}(s_{g1}),\ldots,C_{g}(s_{gM(g)}) \right)' \hspace*{5mm} \text{ and } \hspace*{5mm} \Delta_{g} = \left(\Delta_{g}(s_{g1}),\ldots,\Delta_{g}(s_{gM(g)}) \right)'\; .
\end{equation}
The expression \eqref{4.2} corresponds to the case in \eqref{3.5}, but
without specifying the attribute-dependent forward rate process
$\{f_{gs} \, : \, 0 \le s \le s_{aM(a)} \}$, the corresponding mean
discount function $\bar{D}_{g}(s)$ on $[0, s_{aM(a)}]$ is assumed to
be continuous in $s$ and then it is uniformly approximated by a $p$th
order polynomial;
\begin{equation}
\label{4.3}
\bar{D}_{g}(s) = 1 + ( \delta_{11} z_{1g} + \delta_{12} z_{2g} + \delta_{13} z_{3g} ) s + \cdots + (\delta_{p1} z_{1g} + \delta_{p2} z_{2g} + \delta_{p3} z_{3g}) s^{p} \, , 
\end{equation}
where $z_{1} = 1$, $z_{2} = c_{g}$, and $z_{3} = s_{gM(g)}$ are the
attribute variables of the $g$th bond. In this specification the
parameters are common to all the mean discount functions for $g =
1,\ldots , G$ and hence they are estimable with $G$ bond prices, so
long as $G$ is greater than the number of the parameters
contained. Substituting \eqref{4.3} into \eqref{4.2} yields
\[
\sum_{m=1}^{M(g)} C_{g}(s_{gm}) \bar{D}_{g}(s_{gm}) = a_{g} + \delta_{11} d_{g11} + \delta_{12} d_{g12} + \delta_{13} d_{g13} + \cdots + \delta_{p1} d_{gp1} + \delta_{p2} d_{gp2} + \delta_{p3} d_{gp3} \, ,
\]
where
\[
a_{g} = \sum_{m=1}^{M(g)} C_{g}(s_{gm}) \hspace*{8mm} \text{ and } \hspace*{8mm}  d_{gij}= \sum_{m=1}^{M(g)} C_{g}(s_{gm}) z_{gj} s_{gm}^{j} \; .
\]
Here $i$ in $d_{gij}$ denotes the attribute suffix and $j$ the
polynomial order. Thus letting
\[
x_{g} = (d_{g11}, d_{g21}, d_{g31}; d_{g12}, d_{g22}, d_{g32}; \cdots ; d_{g1p}, d_{g2p}, d_{g3p})'
\]
and
\[
X= (x_{1} , x_{2}, \ldots, x_{G})' 
\]
we have a regression model
\begin{equation}
\label{4.4}
y = X \beta + \eta \, ,
\end{equation}
where $y = (y_{1}, y_{2}, \ldots, y_{G})'$ with $y_{g} = P_{g} -
a_{g}$, $\eta = (\eta_{1}, \ldots, \eta_{G})'$ and 
\[
\beta = (\delta_{11}, \delta_{12}, \delta_{13}; \delta_{21}, \delta_{22},
\delta_{23} ; \ldots ; \delta_{p1}, \delta_{p2}, \delta_{p3})'\, .
\]
In \eqref{4.4} the specification of the covariance matrix $\eta$ is
crucial since specifying the covariance structure of $P = P_{1},
\ldots, P_{G})'$ or equivalently the covariance structure of $\eta$
stochastically describes a structure of the joint realizations of $G$
bond prices. In view of \eqref{4.2} the specification is directly
related to that of the covariances of the stochastic discount factors
$D_{g}(s_{gj})$ and $D_{h}(s_{hm})$ at each cash flow point $s_{gj}$
and $s_{hm}$ of the $g$th and $h$th bonds.  We specify it as
\begin{equation}
\label{4.5}
 Cov(D_{g}(s_{gj}), \, D_{h}(s_{hm})) = \sigma^{2} \lambda_{gh} f_{gh \cdot jm} \, ,
\end{equation}
where $\sigma^{2}$ is a common covariance and covariance factor,
$\lambda_{gh}$ is a covariance part related to the differences of
maturities and $f_{gh \cdot jm}$ is another covariance part related to
the difference of the cash flow points $s_{gj}$ and $s_{hm}$. These
two parts are further specified as
\begin{equation}
\label{4.6}
\lambda_{gh} = 
\begin{cases}
e_{gg} & g=h \\
\rho e_{gh} & g \neq h
\end{cases}
\end{equation}
with
\[
e_{gh} = \exp \left( - \xi | s_{gM(g)} - s_{h M(h)}|\right)
\]
and
\begin{equation}
\label{4.7}
f_{gh \cdot jm} = \exp \left(- \theta |s_{gj} - s_{hm}| \right) \, ,
\end{equation}
where we assume that $0 \le \theta , \, \rho , \, \xi \le 1$. These specifications imply;
\begin{enumerate}
\item[(1)] as is expressed in $e_{gg}$ of $\lambda_{gg}$, the longer
  the maturity of each bond is, the larger the variance of each price
  is,
\item[(2)] as is expressed in $e_{gh}$ of $\lambda_{gh}$, the larger
  the difference of the maturities of two bonds, the smaller the
  covariance is, and
\item[(3)] as is expressed in $f_{gh \cdot jm}$, the closer the two
  cash flow points are, the larger the covariance of the discount
  factors $D_{g}(s_{gj})$ and $D_{h}(s_{hm})$ is.
\end{enumerate}
Under this specification, the covariance matrix of $\eta$ is given by
\begin{equation}
\label{4.8}
Cov(\eta) = (Cov(\eta_{g}, \eta_{h})) = (Cov(P_{g}, P_{h})) = \sigma^{2} (\lambda_{gh} \varphi_{gh}) \equiv \sigma^{2} \Phi(\theta, \rho, \xi)
\end{equation}
with 
\[
\varphi_{gh} = \sum_{j=1}^{M(g)} \sum_{m=1}^{M(m)} C_{g}(s_{gj}) C_{h}(s_{hm}) f_{gh \cdot jm} \; .
\]

As in \citet{kariya:2004}, the unknown parameters are efficiently
estimated by the GLS (generalized least squares) method, in which we
minimize
\begin{equation}
\label{4.9}
\psi(\beta, \theta, \rho) = \left[y - X \beta \right]' \left[ \Phi(\theta, \rho, \xi) \right]^{-1}\left[y - X \beta \right] 
\end{equation}
with respect to the unknown parameters. First, for given $(\theta, \rho, \xi)$, the minimizer of this function with respect to $\beta$ is known to be the GLSE;
\[
\hat{\beta}(\theta, \rho, \xi) = [ X' \Phi(\theta, \rho, \xi)^{-1} X]^{-1} X' \Phi(\theta, \rho, \xi)^{-1} y
\]
and then the marginally minimized function $\psi(\hat{\beta}, \theta,
\rho, \xi)$ with substitution $\hat{\beta}(\theta, \rho, \xi)$ is
minimized with respect to $(\theta, \rho, \xi)$, yielding the GLSE
$(\hat{\beta}, \hat{\theta}, \hat{\rho}, \hat{\xi})$ where a grid
point method for split points of $(\theta, \rho, \xi)$ may be used.

In \citet[][pp 55--63]{kariya:2004}, an empirical performance due to
\citet{kariya:1994} is demonstrated as an example of GLS estimation
where the order of the polynomial in \eqref{4.3} is set to 2 with
$z_{1} = 0$, $z_{2} = \text{coupon}$, and $z_{3} =
\text{maturity}$. Even in this case, the residual standard deviations
are 0.338 yen for December 27, 1989 with $G= 70$ and 0.312 yen for
January 31, 1990 with $G = 70$ where the face value is 100.  Recently,
the effectiveness of this model is empirically and comprehensively
tested with Japanese bond data in \citet{kariya:2011}.

It is noted that in the specification of the attribute-dependent mean
discount function in \eqref{4.3}, setting $z_1 =1$, $z_2 = 0$, and
$z_3 = 0$ yields the attribute-free specification. In this case the
parameters involved are estimated in the same manner to get the
attribute-independent mean discount function $\bar{D}(s)$.  This
$\bar{D}(s)$ is converted to a yield curve by $R_{s} = -s^{-1} \log
\bar{D}(s)$, which is often referred to as a risk-free yield
curve. This attribute-free discount function may be used to price a
CDS in Section 6 because the cash flows in CDS does not reflect the
same cash flow pattern as the corresponding CB.

\section{CB pricing model for deriving TSDBs}

In this section we propose a formulation of the CB pricing model that
enables us to derive the TSDPs and RRs for each pair of industry index
and rating index where (1) CB price data, (2) data on industry-wise
sales ratios of firms that issued the CB’s and (3) credit rating data
are assumed to be given at $t = 0$. As has been stated in Sections 1
and 2, in our modeling the differences of the business line portfolios
of firms are taken into account in terms of sales ratios of the
firms. As a matter of a fact, each firm has different exposures to
industry-wise business factors and so it has a different profit
structure in association with industry business cycles, which is
greatly relevant to analyzing credit qualities of CBs.

To formulate our model, suppose that at $t = 0$ there are $K$ CBs to
analyze and let $\{ s_{kl} ; \, l=1,\ldots, M(k)\}$, $k=1,2,\ldots, K$ with $s_{k1} < s_{k2}<\cdots<s_{kM(k)}$ denote the future cash flow time points of those CB’s as in the case of GBs. Also let $s_{aM(a)}= \max_{k} s_{kM(k)}$ and let the cash flow function $C_{k}(s)$ of the $k$th CB be defined on $0 < s \le s_{aM(a)}$ for all $k$ though it is zero except for the above finite points.

On the other hand, if the firm that issued a CB gets defaulted before
its maturity, the coupons to be paid thereafter are not paid and some
portion of the face value 100 may be paid after a long procedure of
legal and practical settlements, where the portion relative to 100 is
called recovery rate (RR), which is in general of a stochastic nature.
The expected or empirically averaged value of the RR is known to
depend on its credit grade via a rating agency. Hence actual cash
flows from a CB depend on how likely a firm that issued a CB is to get
defaulted and what the expected RR is when it gets defaulted. In the
market the TSDPs are evaluated simultaneously and consistently
together with the expected RRs and are implicitly reflected in their
market prices of CBs.

\subsection{Basic formulation of CB pricing model}

On this viewpoint, let $\tau_{k}$ be the first random time of the $k$th CB to default and let
\begin{equation}
\label{5.1}
L_{ks} = 
\begin{cases}
0 & \tau_{k} > s \\
1 & \tau_{k} \le s \; .
\end{cases}
\end{equation}
Then $L_{ks}$ defined at $t=0$ is the indicator function of default
event $\{\tau_{k} \le s\}$ and identifies if the $k$th CB gets
defaulted before or on a future time point s. Then the actual cash
flow function $\tilde{C}_{k}(s_{kj})$ at a future coupon generating time $s_{kj}$ is expressed as
\begin{equation}
\label{5.2}
\tilde{C}_{k}(s_{kj}) = C_{k}(s_{kj}) (1 - L_{ks_{kj}}) + 100 \gamma(i(k)) L_{k s_{kj}} (1 -  L_{ks_{kj-1}}) \; .
\end{equation}
This means that $\tilde{C}_{k}(s_{kj}) = C_{k}(s_{kj})$ if
$L_{ks_{kj}}=0$, or equivalently if the firm has not defaulted until
$s_{kj}$, and $\tilde{C}_{k}(s_{kj}) =100 \gamma(i(k))$ if $ L_{k
  s_{kj}} (1 - L_{ks_{kj-1}}) = 1$, or equivalently if it had not
defaulted at $s_{kj-1}$ and has defaulted at $s_{kj}$.  Here
$\gamma(i(k))$ is the mean RR of the $k$th bond with rating grade
$i(k)$, where the credit rating is indexed by natural number $i = 1,
2,\ldots , I$ and the smaller the number is, the higher the credit
grade is. Here this specification implicitly assumes that the recovery
payment is made at $s_{kj}$ if a default event occurs in interval $(
s_{kj-1} , s_{kj} ]$. Note that $s_{kj} - s_{kj-1}$ is typically
about a half year.

However, the expression \eqref{5.2} itself exhibits the future
relation and has never been realized at $t = 0$. Therefore investors’
expected cash flow at $t = 0$ is formulated as for interval $(s_{kj-1}, s_{kj}]$ 
\begin{align}
\label{5.3}
\bar{C}_{k}(s_{kij}) & = C_{k}(s_{kj}) [ 1- p_{k}(s_{jk} : i(k))] \\
& \hspace*{10mm} + 100 \gamma(i(k))[ p_{k}(s_{kj} : i(k)) - p_{k}(s_{kj-1} : i(k))] \chi_{k}(s_{kj}) \, , \nonumber
\end{align}
where $p_{k}(s_{jk} : i(k)) = E[ L_{k s_{kj}}]$ is the default
probability corresponding to the event $\{ \tau_{k} \le s_{kj}\}$,
which is the probability that the $k$th CB gets defaulted before or on
$s_{kj}$ and $\chi_{k}(s_{kj})$ is the indicator function of the cash
flow time points $\{s_{kl} ; l = 1,\ldots , M (k )\}$ of the $k$th
CB. With investors’ expected cash flows in \eqref{5.3} we formulate
our CB model as
\begin{equation}
\label{5.4}
V_{k} = \sum_{j=1}^{M(k)} \bar{C}_{k} (s_{kj}) D_{k}(s_{kj}) \; .
\end{equation}

Before we proceed further, it is remarked that a typical model in the
mathematical finance literature is the conditional univariate model
that introduces the credit element into the discount part;
\[
V_{k} = \sum_{j=1}^{M(k)} C_{k} (s_{kj}) E_{0} \left[ \exp\left(- \int_{0}^{s_{kj}} (r_{u} + \lambda_{u}^{k}) \, du \right)\right]\, ,
\]
where $\{r_{u}\}$ is a Markovian spot rate process and
$\{\lambda_{u}^{k}\}$ is an instantaneous Markovian default intensity
process that discounts cash flows together with the spot rate process.
In this expression it is assumed that $\{r_u \}$ is independent of $\{
L_{ks} \}$ , which is unlikely but yields
\[
E_{0} \left[(1-L_{ks} ) \exp\left( - \int_{0}^{s} r_u \, du \right) \right] = E_0 \left[ \exp\left( -\int_{0}^{s} \left( r_u + \lambda_u^k \right)\, du \right) \right]
\]
by the Doob-Meyer Theorem and hence the default intensity $\{
\lambda_u \}$ are assumed to be exogenously independent of $\{ r_u \}$
. It is often the case that such processes as CIR model are assumed
for $\{ \lambda_u \}$ , the discount function $E_0 [~]$ is evaluated
analytically and then the unknown parameters therein are calibrated or
estimated with the present or past data of the $k$th CB only. This
model is sometimes extended to a multivariate case where the
correlation of $\lambda_u^k$ and $\lambda_u^j$ is often assumed to be
constant. Clearly this modeling approach is quite different from ours.

Another remark is that the expression \eqref{5.3} may be regarded as
the conditional expectations of investors and in that way the discount
function in \eqref{5.4} and hence the model \eqref{5.4} itself may be
regarded as a conditional expression to get a dynamic model.  Now
coming back to our case, the expression in \eqref{5.4} corresponds to
the one in \eqref{3.5} and hence the rest of the argument is similar
to the non-defaultable case. That is, the stochastic discount function is decomposed as
\begin{equation}
\label{5.5}
D_{k}(s) = \bar{D}_{k}(s) + \Delta_{k}(s)
\end{equation}
and as $\bar{D}_{k}(s)$ we use the mean discount function estimated
with $G$ government bond prices where the attributes of the $k$th bond
are inserted into the discount function in evaluation. Consequently
\begin{equation}
\label{5.6}
\{ \bar{D}_{k}(s_{kj})\, : \, j=1,\ldots, s_{kM(k)}\, ; \, k=1,\ldots, K\}
\end{equation}
is a set of known values and it follows from \eqref{5.4} and
\eqref{5.5} that the CB pricing model is
\begin{equation}
\label{5.7}
V_k = \sum_{j=1}^{M(k)} \bar{C}_{k} (s_{kj}) \bar{D}_{k}(s_{kj}) + \varepsilon_{k} \hspace*{4mm} \text{ with } \hspace*{4mm}\varepsilon_{k} = \sum_{j=1}^{M(k)} \bar{C}_{k} (s_{kj}) \Delta_{k}(s_{kj}) \; .
\end{equation}
It is remarked that a joint model of GBs and CBs can be formulated in
which the common mean discount functions are estimated simultaneously,
though we do not follow this because of its complexity.  Note that
there are about 3,000 CBs in the Japanese market and about 50,000 CBs
in the US market.

\subsection{Specification of TSDPs with business portfolio structures and credit discounts}

Next we specify the default probability function of the $k$th bond in
\eqref{5.3} as
\begin{equation}
\label{5.8}
P_{k}(s : i(k)) \equiv \sum_{j=1}^{J} w_{k}(j) p(s : i(k), j)\, ,
\end{equation}
where $w_k (j) \ge 0$,  $\sum_{j=1}^{J} w_{k}(j) =1$.  Here
\begin{equation}
\label{5.9}
p(s: i, j) \hspace*{8mm} i=1,\ldots, I , \hspace*{4mm} j=1, \ldots, J
\end{equation}
is the generic or common TSDP with credit grade $i$ and industry $j$ ,
which is independent of specific CBs. In this paper it is assumed to
be approximated by a polynomial of the $q$th order;
\begin{equation}
\label{5.10}
p(s: i, j) = \alpha_1^{ij} s + \alpha_2^{ij} s^{2} + \cdots + \alpha_{q}^{ij} s^{q} \; .
\end{equation}
On the other hand, $\{ w_k (1),\ldots, w_k ( J ) \}$ in \eqref{5.8} is
the set of the sales ratios of the $k$th CB issuer corresponding to
the industry indices $j = 1,\ldots, J$ and it is regarded as a
business portfolio in terms of industry-wise sales, where the industry
categorization is determined in advance.

Now from \eqref{5.8} and \eqref{5.10} the TSDP of the $k$th bond is
expressed as
\begin{align}
\label{5.11}
p_k (s : i(k)) & = s w_k ' \alpha_1^{i\cdot} + s^2 w_k ' \alpha_2^{i\cdot} + \cdots + s^{q} w_k ' \alpha_q^{i\cdot} \\
& = (s w_k ' , s^2 w_k ' , \ldots, s^q w_k') \beta(i) \nonumber \\
& \equiv w_k (s)' \beta(i) \nonumber
\end{align}
where
\begin{align*}
w_k & = (w_k (1), \ldots, w_k (J))', \\ 
 \alpha_h^{i\cdot} & =  ( \alpha_h^{i1}, \ldots,  \alpha_h^{iJ})' \hspace*{5mm} h=1,\ldots q, \\
\beta(i) & \equiv  (\alpha_1^{i\cdot '} , \ldots, \alpha_q^{i\cdot '} )'\\ 
w_{k}(s)' & = (s w_k ', \ldots, s^q w_k ') \; .
\end{align*}
Also the expected cash flow function in \eqref{5.3} is expressed as
\begin{equation}
\label{5.12}
\bar{C}_{k} (s_{kj}) = C_{k}(s_{kj}) + z_{k} (s_{kj}, s_{kj-1} \, : \, \gamma(i(k)))' \beta(i) \, ,
\end{equation}
where
\[
z_{k} (s_{kj}, s_{kj-1} \, : \, \gamma(i(k)))' = - C_{k}(s_{kj}) w(s_{kj})' + 100 \gamma(i(k)) \chi_{k}(s_{kj}) [ w (s_{kj})' - w(s_{kj-1})'] \; .
\]
Therefore the pricing model in \eqref{5.7} is given by
\begin{align*}
V_{k} & = \sum_{j=1}^{M(k)} C_{k}(s_{kj}) \bar{D}_{k}(s_{kj}) + \sum_{j=1}^{M(k)}  \bar{D}_{k}(s_{kj})z_{k}' (s_{kj}, s_{kj-1} \, : \, \gamma(i(k))) \beta((i(k)) + \varepsilon_{k} \\
& = \hat{P}_{k} + [u_k + \gamma(i(k)) v_k]' \beta(i) + \varepsilon_k \, ,
\end{align*}
which yields a regression model
\begin{equation}
\label{5.13}
y_{k}^{i(k)} = [ u_{k} + \gamma(i(k)) v_{k}]' \beta(i) + \varepsilon_{k}
\end{equation}
with
\begin{align*}
y_{k}^{i(k)} & = V_{k} - \hat{P}_{k} \, ,\\
\hat{P}_{k} & = \sum_{j=1}^{M(k)} C_{k}(s_{kj}) \bar{D}_{k}(s_{kj}) \, ,\\
u_{k} & = - \sum_{j=1}^{M(k)} C_{k}(s_{kj}) \bar{D}_{k}(s_{kj}) w_{k}(s_{kj}) \hspace*{4mm} \text{and}\\
v_{k} & = 100 \sum_{j=1}^{M(k)}  \bar{D}_{k}(s_{kj}) \chi_{k}(s_{kj}) [ w_k (s_{kj}) - w_k (s_{kj-1})] \; .
\end{align*}
In this expression, $\hat{P}_{k}$ is regarded as an expected (or a
theoretical) GB price with non-defaultable cash flow $\{C_{k}(s_{kj})
\, : \, j=1, \ldots, M(k) \}$, and hence $y_{k}^{i(k)}$ is the
difference between the $k$th CB price with credit grade $i ( k )$ and
the corresponding non-defaultable bond price. The difference
$y_{k}^{i(k)}$ tends to be non-positive, because
$\bar{C}_{k}(s_{kj}) \le C_{k}(s_{kj})$ and a CB is of the less
creditability and less liquidity than the corresponding GB. We simply
call $y_{k}^{i(k)}$ the credit risk discount of the $k$th CB price.

Consequently \eqref{5.13} forms a regression model for the credit risk
discounts. In fact, supposing that there are $K_{i}$ CBs of credit
grade $i$ and letting $y(i) = (y_{1}^{i}, \ldots, y_{K_{i}}^{i})'$,
$X_{1}(i) = (u_{1}^{i'}, \ldots, u_{K_{i}}^{i'})'$ and $X_{2}(i) =
(v_{1}^{i'}, \ldots, v_{K_{i}}^{i'})'$ the pricing model for CBs of
credit grade $i$ is reduced to a regression model;
\begin{equation}
\label{5.14}
y(i) = X(i, \gamma(i)) \beta(i) + \varepsilon(i)
\end{equation}
with $X(i, \gamma(i)) = X_{1}(i)+ \gamma(i)X_{2}(i)$, $i=1,\ldots, I$.
Here $K_{i} \ge 2 J q$ is assumed for the identifiability and
estimability of $\beta(i)$ and $\gamma(i)$.  In fact, $K_{i} \ge 2 J
q$ is a necessary condition for the uniqueness of $(\beta(i),
\gamma(i))$ in the sense that $ X(i, \gamma(i)) \beta(i) = X(i,
\gamma(i)^*) \beta(i)^*$ implies $\beta(i)= \beta(i)^*$ and $\gamma(i)
= \gamma(i)^*$. It follows from \eqref{5.14} that the credit risk
discount vector $y (i )$ of the $i$th credit grade is explained by the
regression matrix $X( \gamma(i))$ which depends on the unknown RR
$\gamma(i)$. Combining all the regression models over $i = 1,\ldots ,
I$ yields
\begin{equation}
\label{5.15}
y = X \beta + \varepsilon
\end{equation}
with $X \equiv [X(i, \gamma(i))]$ and $[X(i, \gamma(i))]$ is the $K
\times IJq$ block-diagonal matrix with the $i$th block matrix $X(i,
\gamma(i))$ and $\beta = (\beta(1)', \ldots, \beta(I)')'$ with $K=
K_{1} + \cdots + K_{I}$.

\subsection{Specification of covariance structures}

To specify the covariance structure, we write the dependency of the mean cash
flow function on unknown parameters as $\bar{C}_{k}(s_{kj}) \equiv \bar{C}_{k}(s_{kj} \, : \, \beta(i), \gamma(i(k)))$.  Then the error term in \eqref{5.7} is given by
\[
\varepsilon_{k} = \sum_{j=1}^{M(k)} \bar{C}_{k} (s_{kj} \, : \, \beta(i), \gamma(i(k))) \Delta_{k}(s_{kj}) \; .
\]
Hence the covariance of two error terms of the $k$th and $l$th CB
prices with rating $i (k )$ and $i(l )$ respectively is assumed to be

\begin{align}
\label{5.17}
Cov(\varepsilon_{k}, \varepsilon_{l}) & = \sum_{j=1}^{M(k)} \sum_{m=1}^{M(m)} \bar{C}_{k}(s_{kj} \, : \, \beta(i), \, \gamma(i(k))) \bar{C}_{l}(s_{lm} \, : \, \beta(i), \, \gamma(i(l))) Cov(\Delta_{kj}, \Delta_{lm}) \\
& = \sigma^{2} \lambda_{kl} \varphi_{kl}\, , \nonumber 
\end{align}
where
\[
\lambda_{kl} =
\begin{cases}
e_{kk} &  k=l \\
\rho_{ii} e_{kl} & k \neq l , \, i(k)=i(l)=i \\
\rho_{ij} e_{kl} &k \neq l , \, i(k)=i , \, i(l)=j, \, i\neq j 
\end{cases}
\]
with
\[
e_{kl} = \exp\left(- \xi_{ij} \left|s_{kM(k)} - s_{lM(l)} \right| \right) \hspace*{8mm}   i(k)=i , \, i(l)=j,
\]
and
\[
\varphi_{kl} = \sum_{j=1}^{M(k)} \sum_{m=1}^{M(m)} \bar{C}_{k}(s_{kj} \, : \, \beta(i), \, \gamma(i(k))) \bar{C}_{l}(s_{lm} \, : \, \beta(i), \, \gamma(i(l))) b_{kl \cdot jm}
\]
with 
\[
b_{kl \cdot jm} = \exp\left(- \theta \left| s_{kj} - s_{lm} \right| \right) \; .
\]
In this specification, when the $k$th CB and the $l$th CB are of the
same credit rate, i.e., $i(k ) = i(l ) = i$ , then the covariance is
of the same form as the one in the non- defaultable case. If the two
CBs are not in the same credit category, $\rho_{ij}$ will account for
the cross-correlation between the two categories. Using the above
specifications, the covariance matrices for regression models are
obtained;
\begin{align}
\label{5.18}
Cov(\varepsilon(i)) &= (Cov(\varepsilon_{k}, \varepsilon_{l})) = \sigma^{2}(\lambda_{kl} \varphi_{kl}) = \sigma^{2} \Phi(\beta(i), \gamma(i), \rho_{ii}, \xi_{ii}) \equiv  \sigma^{2} \Phi_{ii} \nonumber \\
Cov(\varepsilon(i), \varepsilon(j)) & = \sigma^{2} \Phi(\beta(i), \beta(j), \gamma(i), \gamma(j), \rho_{ij}, \xi_{ij}) \equiv  \sigma^{2} \Phi_{ij} ~~~~ i \neq j \, \, \text{ and }\\
Cov(\varepsilon) & = (Cov(\varepsilon(i), \varepsilon(j))) = \sigma^{2} \{ \Phi_{ij} \} \; .\nonumber \\
\end{align}
In the CB case, the covariance matrices depend on the regression
coefficient $\beta(i)$'s.

\subsection{Estimation procedure}

Under these formulations we propose the following grid procedure to
simplify the model estimation.
\begin{enumerate}
\item[(1)] For each credit category, the GLS estimation is pursued: fix $i$. 
\begin{enumerate}
\item For each given $(\gamma(i), \rho_{ii}, \xi_{ii})$, each of which
  moves over $0, 0.1,\ldots,0.9$, $\beta(i)$ is estimated by a
  repeated procedure. Setting $\beta(i) = 0$ in $\Phi_{ii}$ and
  minimizing
\[
\psi \equiv [ y(i) - X(i, \gamma(i)) \beta(i)]' [\Phi_{ii} (\beta(i), \gamma(i), \rho_{ii}, \xi_{ii})]^{-1}  [y(i) - X(i, \gamma(i)) \beta(i)]
\]
yields the first step GLSE with $ \Phi_{ii}^{(0)} = \Phi_{ii} (0,
\gamma(i), \rho_{ii}, \xi_{ii})$;
\[
\hat{\beta}(i)^{(1)} = [X(i, \gamma(i))'\Phi_{ii}^{(0)^{-1}}X(i, \gamma(i))]^{-1} X(i, \gamma(i))' \Phi_{ii}^{(0)^{-1}} y(i) \; .
\]
Substituting this $\hat{\beta}(i)^{(1)}$ into $\Phi_{ii}$ to get
$\Phi_{ii}^{(1)} = \Phi_{ii} (\beta(i)^{(1)}, \gamma(i), \rho_{ii},
\xi_{ii})$, applying the same procedure yields the second step MLE
$\hat{\beta}(i)^{(2)}$. Repeating this procedure a couple of times
with a performance evaluation rule, we obtain an approximate minimizer
$(\hat{\beta}(i)^{(n)}, \, \Phi_{ii}^{(n-1)})$ for given $(\gamma(i),
\rho_{ii}, \xi_{ii})$.

\item Repeating the procedure in (a) over possible values of
  $(\gamma(i), \rho_{ii}, \xi_{ii})$, we obtain $(\hat{\beta}(i)^* ,
  \Phi_{ii}^* , \gamma(i)^* , \rho_{ii}^* , \xi_{ii}^*)$ that
  minimizes the objective function approximately. Once we obtain this,
  we may repeat (a) in the neighborhood of the approximate optimal
  value by splitting the neighborhood with a finer division.
\end{enumerate}
\item[(2)] A simultaneous estimation procedure for the combined model
  is obtained as follows.  Fix $(\hat{\beta}(i)^*, \hat{\gamma}(i)^* ,
  \hat{\rho}_{ii}^* , \hat{\xi}_{ii}^*)$ $i=1, \ldots, I$ in the
  matrix 
\[
\Phi_{ij} = \Phi(\beta(i), \beta(j), \gamma(i), \gamma(j),
  \rho_{ij}, \xi_{ij}) \hspace*{10mm} i \neq j \, ,
\]
meaning that the full covariance matrix in \eqref{5.18} carries the $I
( I-1)$ unknown parameters $\{\rho_{ij}, \xi_{ij} \, : \, i \neq j
\}$. Hence in the same way as in (1) we minimize
\[
\Psi(\beta, \rho_{ij}, \xi_{i}) = [y-X\beta]' [ \{\Phi_{ij} (\rho_{ij}, \xi_{ij}) \}]^{-1} [y - X\beta]
\]
to obtain the first approximate simultaneous GLSE
$\hat{\beta}_{full}^{[1]}$. One may replace $\hat{\beta}(i)^*$'s in
the fixed $(\hat{\beta}(i)^*, \hat{\gamma}(i)^* , \hat{\rho}_{ii}^* ,
\hat{\xi}_{ii}^*)$ $i=1, \ldots, I$ by $\hat{\beta}_{full}^{[1]}$ and
repeat the minimization procedures again to get the second approximate
simultaneous GLSE $\hat{\beta}_{full}^{[2]}$.
 
\item[(3)] This minimization procedures are repeated over the degrees
  $q = 2,\ldots,6$ of the polynomials in \eqref{5.10} to get an
  optimal simultaneous GLSE $\hat{\beta}_{full}$.
\end{enumerate}
With this model, an extensive empirical research is under a way with
Japanese CB data. A simple empirical work is demonstrated below in a
limited model. In this model we do not take into account the fact that
each firm has a specific business portfolio over industries. In other
words, we set $J = 1$ and pick the AA class of 208 CBs with the
assumption of recovery rate $\gamma=0$. A plot (see Figure 1) of the residuals
$\{\hat{\varepsilon}_{k}^{AA} = V_{k} - \hat{V}_{k} \}$ shows that
even in this simplified model the performance are rather good and hence
our full model will be more promising as a model to derive the
TSDPs. In fact, the standard deviation of
$\hat{\varepsilon}_{k}^{AA}$'s is $0.419$ yen.

\begin{figure}[H]
\begin{centering}
\includegraphics[height=7in,width=5in]{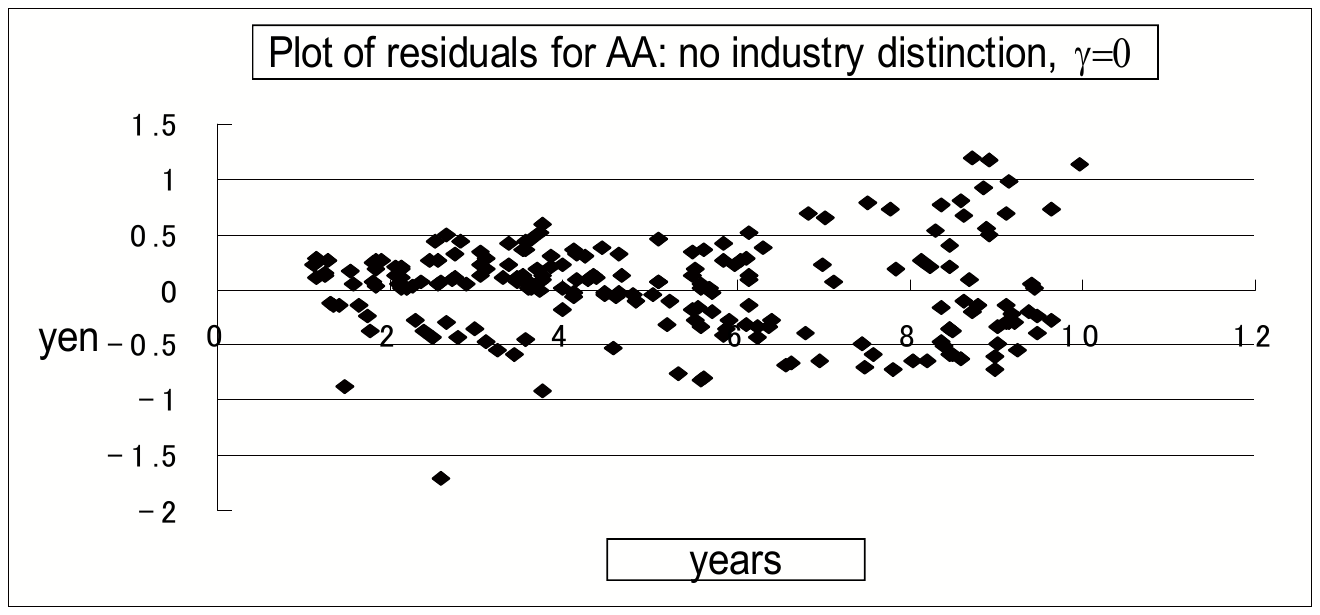}
\par\end{centering}
\end{figure}

\FloatBarrier

\section{Applications}

Once a full estimator $\hat{\beta}_{full}$ is obtained, the following applications are considered.

[1] The whole implied TSDPs in \eqref{5.10} for each pair of
  credit grade index and industry index are given by
\begin{equation}
\label{6.1}
\hat{p}(s : i, j) = \hat{\alpha}_{1}^{ij} s + \hat{\alpha}_{2}^{ij} s^{2} + \cdots + \hat{\alpha}_{q}^{ij} s^{q}~~~i=1,\ldots, I~~j=1,\ldots, J
\end{equation}
together with the implied RRs $\{\hat{\gamma}(i)^* : i = 1, \ldots, I
\}$. These TSDPs and RRs are basic information sources for
credit-related products.

[2] The TSDP of the $k$th CB or its issuer with credit grade $i (k )$
and business portfolio $\{w_{k}(1), \ldots, w_{k}(J) \}$ over $J$
industries is obtained from [1] as
\begin{equation}
\label{eq:6.1}
\hat{p}_{k}(s : i(k)) = \sum_{j=1}^{J} w_{k}(j) \hat{p}(s : i(k), j)\; .
\end{equation}
This is the TSDP $\{P(\tau_{k} \le s) : s > 0 \}$ 0f the issuer of the
$k$th CB where $\tau_{k}$ is the default time as in \eqref{5.1}. The
expected RR is $\hat{\gamma}(i(k))^* \equiv \hat{\gamma}(i)$ with $i(k)=i$.

[3] Inserting $\hat{\beta}_{full}$ into \eqref{5.13} and using
\eqref{5.12}, the credit risk discount level of the $k$th CB or its
issuer with credit grade $i (k )$ and business portfolio $\{w_{k}(1),
\ldots, w_{k}(J) \}$ over $J$ industries is given by 
\begin{align}
\label{6.3}
\hat{y}_{k}^{i(k)} & = [ u_{k} + \hat{\gamma}(i(k))^* v_{k} ]' \hat{\beta}_{full} \\
& = \sum_{j=1}^{M(k)} \bar{D}_{k} (s_{kj}) W(\hat{p}_{k}(s_{kj} : i(k)), \hat{p}_{k}(s_{kj-1} : i(k)); \hat{\gamma}(i(k))) \nonumber
\end{align}
with
\begin{align}
\label{6.3a}
\hat{W}(s_{kj}) & \equiv  W(\hat{p}_{k}(s_{kj} : i(k)), \hat{p}_{k}(s_{kj-1} : i(k)); \hat{\gamma}(i(k))) \\
 & = [ 100 \hat{\gamma}(i(k)) \chi_{k}(s_{kj}) - C_{k}(s_{kj})] \hat{p}_{k}(s_{kj} : i(k)) - 100 \hat{\gamma}(i(k)) \nonumber \\
 & \hspace*{10mm} \times \hat{p}_{k}(s_{kj-1} : i(k))\chi_{k}(s_{kj}) \nonumber
\end{align}
where the discount is relative to the fair value $\hat{P}_{k} =
\hat{V}_{k} + (-\hat{y}_{k}^{i(k)})$ of the corresponding
non-defaultable bond. Here $\hat{P}_{k}$ is the model value of
non-defaultable bond with the same coupon and same maturity as those
of the $k$th CB. Consequently the fair spread rate of the $k$th CB
relative to the corresponding (attribute-dependent) GB is given by
\begin{equation}
\label{6.4}
s \hat{p}_{k} \equiv ( \hat{P}_{k} - \hat{V}_{k}) /  \hat{P}_{k} = -\hat{y}_{k}^{i(k)} / \hat{P}_{k} \; .
\end{equation}
Hence if the actual spread $(\hat{P}_{k} - \hat{V}_{k}) / \hat{P}_{k} =
-\hat{y}_{k}^{i(k)} / \hat{P}_{k}$ is too low or too high relative to
this fair spread, one may use the information for investment
decision making with the expectation that the actual spread eventually
converges to the fair level.  

[4] CB Portfolio Management. The fair value of a portfolio $\{ \varpi_{k}\}$ of CBs is given by
\begin{align}
\label{6.5}
\sum_{k=1}^{K} \varpi_{k} \hat{V}_{k} & = \sum_{k=1}^{K} \varpi_{k} \sum_{m=1}^{M(a)} \hat{\bar{C}}_{k} (s_{am}) \bar{D}_{k} (s_{am}) = \sum_{k=1}^{K} \varpi_{k} [ \hat{P}_{k} + \hat{y}_{k}] \\
& = \sum_{m=1}^{M(a)} [ A_{am} + B_{am}] \equiv  \sum_{m=1}^{M(a)}  C_{am}\nonumber 
\end{align}
where $\{ \varpi_{k}\}$ is a portfolio of CBs in terms of physical units,
\begin{equation}
\label{6.5a}
A_{am} = \sum_{k=1}^{K} \varpi_{k} C_{k} (s_{am}) \bar{D}_{k} (s_{am})
\end{equation}
and
\begin{equation}
\label{6.5b}
B_{am} = \sum_{k=1}^{K} \varpi_{k}  \bar{D}_{k} (s_{am}) W(\hat{p}_{k}(s_{am} : i(k)), \hat{p}_{k}(s_{am-1} : i(k)); \hat{\gamma}(i(k))) \; . 
\end{equation}
Here $A_{am}$ is the present value of the cash flows at future time
$s_{am}$ if none of the bonds are defaulted before or on $s_{am}$
where $\{s_{am}\}$ is the combined set of cash flow points, while
$B_{am}$ is the present value of the expected loss at $s_{am}$, which
is generally negative.

What is important in this expression is that these are regarded as
fair present values of those of the corresponding loan portfolio in a
bank since a fixed-rate loan is regarded as a CB and a variable-rate
loan may be converted to an equivalent fixed loan via interest swap
rate at the time of evaluation. Though we do not elaborate it, based
on this expression, the portfolio can be adjusted to a desirable loss
structure relative to industries and rating grades via
\[
\hat{p}_{k}(s ; i) \equiv \sum_{j=1}^{J} w_{k}(j) \hat{p}(s: i, j) \, ,
\]
where \eqref{6.3a}, \eqref{6.5} and \eqref{6.5b} are combined to sort
out the overlapping structure of firms’ portfolios and extract the
industry positions for each $ \hat{p}(s: i, j)$.

Another example is about the durations of expected losses. Putting $b_{m} = B_{am} / B$ where 
\[
B= \sum_{m=1}^{M(a)} B_{am} \, ,
\]
$\{b_{m}\}$ represents the relative sizes of losses distributed over $\{ s_{am}\}$ and hence 
\[
\sum_{m=1}^{M} b_{m} s_{am}
\]
is the average duration of the losses. On the other hand, $\{ a_{m}\}$ with
$a_{m} = A_{am} / A$ and
\[
A = \sum_{m=1}^{M} A_{am} \, ,
\]
represents the relative sizes of the cash inflows distributed over $\{
s_{am} \}$ if none of the CBs or loans get defaulted, and so the
duration is
\[
\sum_{m=1}^{M} a_{m} s_{am} \; .
\]
The two durations are in general different and the expected actual
duration becomes
\[
\sum_{m=1}^{M} c_{m} s_{am} 
\]
with $c_{m}= C_{am} /C$ and
\[
C = \sum_{m=1}^{M} C_{am} \, ,
\]
The above argument can be extended to the case where the portfolio
contains GBs as well as CBs.

[5] CDS Pricing.  Now we apply the result to the valuation of credit
default swap (CDS). The CDS is notorious as a credit derivative for
enlarging and broadening the financial crisis because the protection
sellers like AIG sold CDSs more than their capacities. However CDS
itself is an important instrument because it is an insurance-type
product (contract) guaranteeing the principal of a specific CB (or
loan) for the protection buyers by covering the loss when the issuer C
of the CB gets defaulted.  A protection (insurance) buyer B in this
contract is the one who pays the protection premiums quarterly or
biannually to a protection seller A over a certain period till its
contract period or default of C in the period in order to protect the
principal. The CDSs of this type are quoted in the market for trades
of credit risk on GBs and CBs. The quotation includes the ask-side as
well as the offer-side of protection against each bond, from which a
market value of the credit risk of a CB is found. Here we first give a
price (premium) formula for CDS in a discrete time setting, which is
different from those in a continuous setting. As is stated in Section
1, a discrete time setting is more suitable in credit risk analysis
than a continuous time setting. For example, a credit variation
process is not Markovian and a settlement process after a default
takes often 3 months through 6 months. In the discrete-time setting, the
uncertainty of the time delay for settlement is directly incorporated
into the model if necessary.

Now to price a CDS, let $m= 0, 1, \ldots M$ be daily counted time
points at 0 where the time unit $h$ of day is measured in year as $h =
1/365$. Time point $m$ corresponds to $mh$ years from 0. Let
\begin{equation}
\label{6.6}
\Upsilon = \{0 \le m_{1} \le m_{2} < \cdots < m_{k}  \}
\end{equation}
be the biannual premium payment time points counted in days viewed at
0, and at these time points the protection buyer pays premium $x$ for
principal value 100.  Further let $N^{C}$ be the default time of firm
C as in \eqref{5.1} and let the default process be denoted by
\[
\{ L_{m} \} ~~~~~ \text{ with } ~~~~~ L_{m} = \chi_{\{N^{C} \le m \}} \; .
\]
To derive a theoretical premium $x$ , we distinguish the two cases; $m \notin \Upsilon$ or $m \in \Upsilon$.  

When $m \notin \Upsilon$ the cash flow at $m$ of the protection buyer B
occurs from the protection seller A only if firm C gets defaulted at
$m$. Hence the payoff of A is expressed as
\begin{equation}
\label{6.7}
U_{m}^{A} = -100 (1 - L_{m-1}) L_{m} + 100 \gamma_{m}(1 - L_{m-1}) L_{m}\; .
\end{equation}
In \eqref{6.7}, if C had defaulted at $m - 1$, $L_{m - 1} = 1$ and so
$U_{m}^A = 0$ , i.e., no cash flow at $m$.  Here this expression assumes
that on the day of default, protection seller A pays 100 and receives
recovery $100 \gamma_m$ , which is not realistic, but we can change it
in a way as we wish. In fact, if the protection seller A pays the
principal to B in 14 days after default and A gets the recovery money
in 90 days after default, the expression can be modified as for the
payoff of A at $m$
\begin{equation}
\label{6.8}
U_{m}^{A} = -100 (1 - L_{m-1 - 14}) L_{m - 14} ~~ \text{ and } ~~  U_{m + 90}^{A} = 100 \gamma_{m+90}(1 - L_{m-1-14}) L_{m-14}\; .
\end{equation}
However, for simplicity we here assume \eqref{6.7} in which the
payments are made at $m$ when the default occurs in $(m - 1, m]$.

When $m \in \Upsilon$, B pays $100 x$ to A if the default has not
occurred until $m$ and hence combining this with \eqref{6.7} the
payoff of A becomes
\begin{equation}
\label{6.9}
U_{m}^{A} = -100 (1 - \gamma_{m}) (1 - L_{m-1}) L_{m} + 100 x(1 - L_{m})\; .
\end{equation}
Note that by the definition of $L_{m}$, for $m < n$, $L_{m}=1$ implies
$L_{n}=1$ and hence $U_{m}^A =0$.  Now to value a theoretical premium
$x$ in a way as in no-arbitrage theory, each payoff $U_{m}$ at $m$ is
regarded as a derivative and is valued relative to cash managed up to
$m$ with daily spot rate process $\{r_j\}$;
\begin{equation}
\label{6.10}
B_{m} = \exp \left( \sum_{j=0}^{m-1} r_{j} h \right) \; .
\end{equation}
Then the martingale condition gives the value at 0 of $U_n$ by
\begin{equation}
\label{6.11}
\nu^A (m) = E_0^* [ U_m / B_m] = E_0^* [ d(m) U_m] \, ,
\end{equation}
where in the theory the measure is risk neutral though it is not
uniquely identifiable. If the interest rate process is independent of
$U_n$, which would not hold in reality, the value is expressed as
\begin{equation}
\label{6.12}
\nu^A (m) = 100 \bar{D}(m) [ - (1-\hat{\gamma}(i(C)))Q(N^C = m) + \delta(m) x Q(N^C > m)] \, ,
\end{equation}
where $\delta(m)=1$ if $m \in \Upsilon$, and it is 0 otherwise. Here we use an attribute-free discount function;
\[
\bar{D}(m) = E_0 [ 1 / B_m ] = E_0 [ d(m) ] \; .
\]
This discount function can be chosen as the one stated at the end of Section 4.
Therefore the expected payoff at 0 of A is
\[
V^A = \sum_{m=1}^{M} \nu^A (m) \; .
\]
On the other hand, the payoff of B should be fair to that of A,
meaning that it is equal to $V^A$ and $V^B = -V^A$. Consequently the
premium $x$ must satisfy $V^A = 0$.

The CB-CDS premium described above is valued as a fair value;
\begin{equation}
\label{6.13}
x =  \frac{(1-\gamma(i(C))) \sum_{m=1}^{M} \bar{D}(m) Q(N^C = m)}{\sum_{k=1}^{K} Q(N^C > m_k)  \bar{D}(m_k)} \, ,
\end{equation}
where $\gamma(i(C))$ is the recovery rate of the CB issuer C and $\{
m_{k}\}$ is the set of premium payment times. In our case,
$\gamma(i(C))$ is the implied recovery rate and the default
probability is the implied TSDP,
\[
Q(N^C \le m) = p_C (mh : i(C)) = \sum_{j=1}^{J} w_k (C) p(mh : i(C), j) \; , 
\]
where the industry-wise business structure of the issuer C is taken
into account. In \eqref{6.3}, one may replace
\[
\sum_{m=1}^{M} Q(N^C = m) \bar{D}_C (m)
\]
by the continuous time expression
\[
\int_0^{S_{M(C)}} \bar{D}(s) p_{C} ' (s: i(C)) \, ds \; ,
\]
where $p_{C}' (s: i(C))$ is the derivative of the TSDP $p_C$.

\section{Conclusion}
In this paper, we formulated a CB pricing model from which the TSDPs
are derived for each industry index and each rating index and the
recovery rates are derived for each rating index. A notable feature of
this model is to take into account the fact that each firm has a
different portfolio structure of business lines over industries. The
TSDPs and recovery rates are basic inputs in credit derivatives and
credit risk management.  Though we wait for a thorough empirical work,
the model itself and derived outputs will be very useful for practical
decision makings. Further these results can be also applied to pricing
some multi-name credit derivatives.

\bibliographystyle{imsart-nameyear}
\bibliography{ref}
\end{document}